\documentclass[%
 reprint,
superscriptaddress,
 amsmath,amssymb,
 aps,
]{revtex4-1}

\usepackage[dvipdfmx]{graphicx}
\usepackage{dcolumn}
\usepackage{bm}

\usepackage{color}

\begin{document}

\preprint{APS/123-QED}

\title{Quadratic unconstrained binary optimization formulation\\
 for rectified-linear-unit-type functions}

\author{Go Sato}
\affiliation{%
Graduate School of Science and Enginnering, Saitama University,
255 Shimo-Okubo, Sakura-ku, Saitama-shi, 338-8570, Japan
}%

\author{Makiko Konoshima}
\affiliation{%
Fujitsu Laboratories Ltd.,
4-1-1 Kawasaki, Kanagawa 211-8558, Japan
}%

\author{Takuya Ohwa}
\affiliation{%
Fujitsu Laboratories Ltd.,
4-1-1 Kawasaki, Kanagawa 211-8558, Japan
}%

\author{Hirotaka Tamura}
\affiliation{%
Fujitsu Laboratories Ltd.,
4-1-1 Kawasaki, Kanagawa 211-8558, Japan
}%

\author{Jun Ohkubo}%
 \email{johkubo@mail.saitama-u.ac.jp}
\affiliation{%
Graduate School of Science and Enginnering, Saitama University,
255 Shimo-Okubo, Sakura-ku, Saitama-shi, 338-8570, Japan
}%
\affiliation{%
JST, PREST, 4-1-8 Honcho, Kawaguchi, Saitama, 332-0012, Japan
}%

\date{\today}

\begin{abstract}
We propose a quadratic unconstrained binary optimization (QUBO) formulation of rectified linear unit (ReLU) type functions. Different from the $q$-loss function proposed by Denchev \textit{et al.} (2012), a simple discussion based on the Legendre duality is not sufficient to obtain the QUBO formulation of the ReLU-type functions. In addition to the Legendre duality, we employ the Wolfe duality, and the QUBO formulation of the ReLU-type is derived. The QUBO formulation is available in Ising-type annealing methods, including quantum annealing machines.
\end{abstract}

\pacs{Valid PACS appear here}
\maketitle


\section{Introduction}
Optimization methods are attracting attention in machine learning field. The optimization methods are used to minimize cost functions in machine learning problems. Therefore, the performance of the optimization methods often has a great influence on the results of machine learning. Although, in general, it is difficult to find exact solutions for the optimization problems, approximate solutions could give reasonable results for the purpose of machine learning. Hence, some heuristic methods are often employed to obtain approximate solutions within realistic times.

Here, we focus on combinatorial optimization problems. Although there are various studies to solve the combinatorial optimization problems approximately, an annealing method is one of the famous methods for efficiently solving them. There are two famous annealing concepts \cite{Kirkpatrick, Ray}: one is the simulated annealing method in which the temperature of the system is controlled to search the global minimum; another is the quantum annealing method which uses quantum effects. Especially, the quantum annealing has attracted attention. The current mainstream quantum annealing method was proposed by Kadowaki and Nishimori \cite{Kadowaki}. After that, Farhi \textit{et al.} proposed a similar idea, which was called the adiabatic quantum computing (AQC), and its analysis has been done from the viewpoint of the computational complexity \cite{Farhi}.

In order to employ the quantum annealing ideas, it is important to reformulate an optimization problem as a quadratic unconstrained binary optimization (QUBO) formulation \cite{Boros, Borosetal}. The QUBO formulation is equivalent to Ising models, which are familiar to physicists. Recently, the Canadian company D-Wave has released ``D-Wave 2000'' which implements the AQC \cite{D-Wave}, and the hardware needs the QUBO formulation. Additionally, ``FUJITSU Quantum-inspired Computing Digital Annealer,'' which enables us to perform the rapid annealing computation \cite{Fujitsu}, needs also the QUBO formulation to perform the optimization.

As for the QUBO formulation, there are many researches (See the review paper \cite{Biamonte} for famous examples.) For example, Whitfield \textit{et al.} made it possible to use logic elements for quantum annealing \cite{Whitfield}. Nazareth and Spaans solved the problem of Intensity Modulated Radiation Therapy (IMRT) using quantum annealing and compared it with the traditional method \cite{Nazareth2015}. In the context of machine learning, Denchev \textit{et al.} proposed the $q$-loss function which is robust against the label noise in machine learning, and the QUBO formulation of the $q$-loss function was derived \cite{Denchev}. 
The Legendre transformation has been used to derive the QUBO formulation, and the usefulness of the $q$-loss function has been shown. Therefore, one might expect that the derivation technique will be applicable for other types of functions. However, this is not true; the $q$-loss function has a quadratic part, and the quadratic property is important to derive the QUBO formulation via the Legendre transformation; if we replace the quadratic part with a linear function, the derivation is not straightforward, as will be described later. 
An example of functions with such a linear part is the famous rectified linear unit (ReLU) \cite{Glorot}; the ReLU function has such a linear part in the functional form, and its important role in the deep learning has been well-known in machine learning research community. 
Although the ReLU function has been mainly used in deep neural networks, the ReLU-type functions can be used as penalty terms with other cost functions. In the IMRT example \cite{Nazareth2015}, a penalty term has a constant part and a quadratic part, which is similar to the ReLU-type functions (the ReLU-type function has a constant part and a linear part.) In this sense, apart from the neural network contexts, the ReLU-type functions would be useful for some optimization problems.
However, as far as we know, the QUBO formulation of the ReLU-type functions has not been derived yet.

In the present paper, a QUBO formulation for a ReLU-type function is proposed. To derive the QUBO formulation of the ReLU-type function, we use not only the Legendre transformation, but also the Wolfe dual theorem, which has not been used in the derivation of the QUBO formulation for the $q$-loss function in \cite{Denchev}.

\section{Previous research}
Since the QUBO formulation and the Ising model are equivalent, one could express the QUBO formulation as a two-body Ising model. The two-body Ising model is represented as follows:
\begin{align}
\label{eq:two-body-ising}
\mathcal{H}=-\sum_{i,j}J_{ij}\sigma_{i}\sigma_{j}-\sum_{i}h_{i}\sigma_{i},
\end{align}
where $\sigma_{i}\in\{-1,1\}$ is a spin variable for $i$-th spin, $J_{ij}\in\mathbb{R}$ a two-body interaction between spin $i$ and $j$, and $h_{i}\in\mathbb{R}$ an external field interacting with spin $i$.
Of course, it is straightforward to convert the spin variables to conventional binary variables $\{0,1\}$.

The two-body Ising model has the role of a cost function in the annealing method. That is, as denoted in Introduction, if the cost function of the problem can be represented by a two-body Ising model, the problem can be solved using the annealing method.

\begin{figure}[bt]
  \begin{center}
    \includegraphics{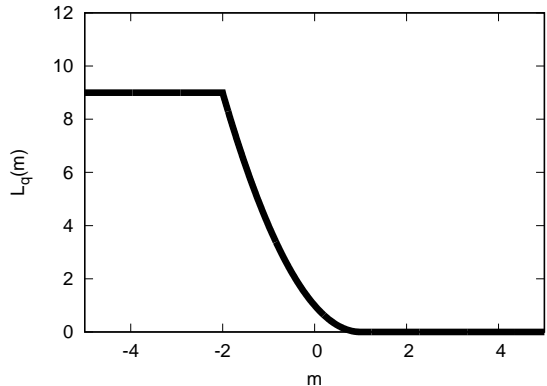}
    \caption{The $q$-loss function with $q = -2$ case.}
  \label{fig:q-loss}
  \end{center}
\end{figure}

The $q$-loss function proposed by Denchev \textit{et al.} is expressed as follows \cite{Denchev}:
\begin{align}
\label{eq:q-loss_formal}
L_{q}(m)=\min\left(\left(1-q\right)^{2},\left(\max(0,1-m)\right)^{2}\right),
\end{align}
where $q\in [-\infty,0)$ is a parameter, which determines the shape of the quadratic part. A specific form of the $q$-loss function with $q = -2$ is shown in Fig.~\ref{fig:q-loss}. In \cite{Denchev}, it has been shown that the $q$-loss function has robust features against label noise; if the input $m$ takes a large negative value, the $q$-loss function takes a constant value, which enables us to obtain a robust classifier against outliers. In order to use the $q$-loss function in the annealing method, the following two-body Ising expression has been derived in \cite{Denchev}:
\begin{align}
\label{eq:q-loss_Legform}
L_{q}(m)=\min_{t}\left\{(m-t)^{2}+(1-q)\frac{1-\mathrm{sign}(t-1)}{2}\right\},
\end{align}
where $t \in \mathbb{R}$ is a new variable introduced through the Legendre transformation. As for details of the derivation of \eqref{eq:q-loss_Legform}, see \cite{Denchev}. 

Note that \eqref{eq:q-loss_Legform} is quadratic for $m$ and $t$ respectively, and is expressed in a form of minimization. Although the variables $m$ and $t$ are continuous, it is straightforward to obtain the QUBO formulation by performing the binary expansions of these continuous variables. The sign function in \eqref{eq:q-loss_Legform} is also expressed as a one-body term when we employ the binary expansion. Furthermore, it is possible to combine the $q$-loss function with other cost functions for the variable $m$ because the $q$-loss function is formulated as the minimization problem. From these reasons, it is clear that the minimization formulation up to second-order (and sign functions) is enough to derive the QUBO formulation.

\section{QUBO formulation for ReLU-type function}

\begin{figure}[bt]
  \begin{center}
    \includegraphics[scale =1.0]{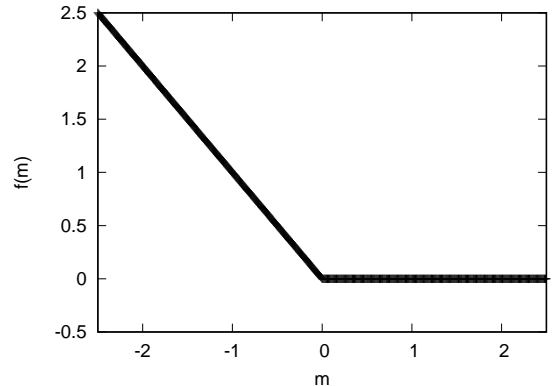}
    \caption{The functional form of $f(m)$.}
  \label{fig:Relu}
  \end{center}
\end{figure}

In this section, we give the main result of the present paper; the QUBO formulation for a ReLU-type function is derived. Similar to the $q$-loss function, we define the following function $f(m)$:
\begin{align}
\label{eq:Relu_min}
f(m) = -\min(0,m).
\end{align}
A functional form of $f(m)$ is shown in Fig.~\ref{fig:Relu}. Note that the function $f(m)$ is converted to the conventional ReLU function by employing the variable transformation $m \to -m$.
Since the ReLU-type function does not have a constant penalty part, the robustness features against label noise is lost, which is different from the $q$-loss function. However, the main purpose of this paper is to derive the QUBO formulation of a function in which the quadratic part in $q$-loss function is replaced with a linear part. Therefore, the robustness feature is ignored in the present paper.

Below, we reformulate the function $f(m)$ in \eqref{eq:Relu_min} as a function up to second order terms. Different from the derivation for the $q$-loss function in \cite{Denchev}, a simple usage of the Legendre transformation is not enough to obtain the adequate functional form for the minimization problem in the QUBO formulation; we will see it soon. Hence, in addition to the Legendre transformation, we here employ the Wolfe dual theorem.

Note that if a function $f_{L}$ is convex, the Legendre transformation of $f_{L}$, the so-called conjugate function of $f_{L}$, is given as follows:
\begin{align}
\label{eq:Legmax}
f_{L}^{*}(t)=\sup_{x}\{t x - f_{L}(x)\}.
\end{align}
In addition, \eqref{eq:Legmax} is equivalent to following equation:
\begin{align}
\label{eq:Legmin}
f_{L}^{*}(t)=-\inf_{x}\{f_{L}(x) - t x\}.
\end{align}

Here, we try naive application of the Legendre transformation to the function $f(m)$ in \eqref{eq:Relu_min}. In order to employ the above definition of the Legendre transformation, we give a slightly different form of \eqref{eq:Relu_min} as follows:
\begin{align}
\label{eq:fmadd}
f(m)= \begin{cases}
-m & (m<0), \\
0 & (m \geq 0).
\end{cases}
\end{align}
Then, the Legendre transformation is performed for each domain as follows:
\begin{itemize}
\item[(a)] $m<0$:\\
The gradient in this domain is always $-1$. Hence,
\begin{align*}
-\inf_m \left\{- m -mt \right\} = -\inf_m \left\{-m(1+t)\right\} = 0,
\end{align*}
and the conjugate function is $f^{*}(t)=0$. Note that the possible value of $t$ is only $t=-1$. 
\item[(b)] $m=0$:\\
Since the left differentiation at this point is $f_{-}'(m)=-1$ and the right differentiation is $f_{+}'(m)=0$, the gradient value takes an arbitrary value within $-1$ to $0$. Hence, the conjugate function is $f^{*}(t)=0$ with the domain $t\in[-1,0]$.
\item[(c)] $0<m$:\\
The gradient in this domain is always $0$. Hence,
\begin{align*}
-\inf_m \left\{-mt \right\} = 0,
\end{align*}
and the conjugate function is $f^{*}(t)=0$. Note that the possible value of $t$ is only $t=0$.
\end{itemize}
From the above discussion, the conjugate function of $f(m)$ is $f^{*}(t) =0\mathrm{\ }(-1\leq t \leq 0)$. When we apply the Legendre transformation to $f^{*}(t)$ again, $f(m)$ is adequately recovered. Therefore, we could find the quadratic form of $f(m)$ as follows:
\begin{align}
\label{eq:same_Lp}
F(m)=-\min_{t}\left\{ -mt \right\}
\quad \mathrm{subject\ to}\quad -1\leq t \leq 0.
\end{align}
In order to emphasize the fact that it is the quadratic form of $f(m)$, we newly introduced $F(m)$ instead of $f(m)$.

Although we could find the quadratic form of $f(m)$ in \eqref{eq:same_Lp}, there is a critical problem for applications. Since there is the minus sign before the $\min$ function in \eqref{eq:same_Lp}, it becomes difficult to solve the optimization problem combining $F(m)$ and another cost function $C(m)$.
That is, although all variables are simultaneously optimized in the annealing, the derived quadratic form $F(m)$ is not suitable for the simultaneous optimization: 
\begin{align*}
\min_{m}\left\{ C(m) + F(m) \right\}  &=  \min_{m} \left\{ C(m) - \min_{t} \{-mt\} \right\} \\
&\neq   \min_{m, t} \left\{ C(m) - (-mt)\right\}.
\end{align*}
Hence, it is not in the form of minimization problem for both $m$ and $t$.

Note that the use of the Legendre transformation to a simple \textit{concave} function avoids this problem; actually, a term, $- m^2$, is added to the $q$-loss function in \cite{Denchev}, and a simple concave function with a linear part is obtained. That is, the $q$-loss function has a quadratic part in the middle region, and the addition of $-m^2$ makes the middle region to a linear part. In contrast, if we have a linear part in the middle region in the $q$-loss function, one can see that a simple usage of the Legendre transformation does not give the QUBO formulation. We also confirmed that a simple conversion of the ReLU-type function into a concave one does not work well.

In order to overcome this problem, we here employ the Wolfe dual theorem \cite{PWolfe}. 
The Wolfe dual theorem is used in nonlinear programming and mathematical optimization. Here, we give a brief explanation for the Wolfe dual theorem. The Wolfe dual theorem is available to derive a dual problem for minimization problem with inequality constraints (main problem). Suppose that an objective function and constraints can be differentiated, and for example, consider the following optimization problem with constraints:
\begin{align}
\label{eq:Wolfetarget}
\left\{
    \begin{array}{lll}
      \mathrm{minimize}_{\bm{x}} & f_{\mathrm{W}}(\bm{x}) & (\bm{x}\in \mathbb{R}^{n}), \\
      \mathrm{subject\ to} & h_{i}(\bm{x}) \leq 0 & (i=1,2,\dots,l),
    \end{array}
  \right.
\end{align}
where $f_{\mathrm{W}}(\bm{x})$ is a certain convex function to be optimized, and $h_i$ are convex and inequality constraints. The Lagrangian function for this optimization problem is
\begin{align}
L(\bm{x},\bm{z})=f_{\mathrm{W}}(\bm{x})+z^T h(\bm{x}),
\end{align}
where $\bm{z}$ is a vector of the Legendre coefficients. Then, the Wolfe dual theorem means that the minimization problem in \eqref{eq:Wolfetarget} is equivalent to the following maximization problem:
\begin{align}
\left\{
    \begin{array}{lll}
      \mathrm{maximize}_{\bm{x},\bm{z}} & L(\bm{x},\bm{z}) & ((\bm{x},\bm{z})\in \mathbb{R}^{n}\times\mathbb{R}^{l}), \\
      \mathrm{subject\ to} & \nabla L(\bm{x},\bm{z})=0 & (\bm{z} \geq 0).
    \end{array}
  \right.
\end{align}

As shown above, the Wolfe dual theorem transforms the minimization problem to the maximization problem. Hence, we here apply it to $F(m)$ in \eqref{eq:same_Lp}. Rewriting the optimization problem of \eqref{eq:same_Lp} in the same form as \eqref{eq:Wolfetarget}, we have
\begin{align}
\label{eq:formoptmin}
\left\{
    \begin{array}{ll}
      \mathrm{minimize}_{t} &-mt,\\
      \mathrm{subject\ to} & -1 \leq t \leq 0.
    \end{array}
  \right.
\end{align}
Then, using the Wolfe dual theorem, we obtain the dual problem of the optimization problem \eqref{eq:formoptmin} as follows.
\begin{align}
\label{eq:wolfeleg}
\left\{
    \begin{array}{ll}
      \mathrm{maximize}_{t,z} &  -mt - z_{1}(t+1) + z_{2} t, \\
      \mathrm{subject\ to} 
      & -m-z_{1}+z_{2}=0, \\
      &-1 \leq t \leq 0,z_{1}\geq 0,z_{2}\geq 0.
    \end{array}
  \right.
\end{align}

It is straightforward to derive the QUBO formulation from the optimization problem \eqref{eq:wolfeleg}. In order to remove the first constraint condition, $-m-z_{1}+z_{2} = 0$, in \eqref{eq:wolfeleg}, it is enough to add the following penalty term to the maximize expression:
\begin{align}
\label{eq:penalty1}
-M\left(-m-z_{1}+z_{2}\right)^2,
\end{align}
where $M$ is a constant and it should take a large value to ensure the equality constraint to be satisfied. The remaining inequality constraint conditions, $ -1 \leq t \leq 0 $, $ z_{1} \geq 0 $ and $ z_{2} \geq 0 $, can be easily realized by expanding these variables, $t$, $z_1$, and $z_2$, in the binary expressions which satisfy the corresponding domain constraints.

From the above, the optimization problem \eqref{eq:wolfeleg} is expressed in the two-body Ising form as follows:
\begin{align}
\label{eq:Isingfunc}
&L(m,t,z_{1},z_{2})\nonumber\\
& = -mt - z_{1}(t+1) + z_{2} t -M\left(-m-z_{1}+z_{2}\right)^2.
\end{align}
By using $L(m,t,z_{1},z_{2})$, the following relationship holds.
\begin{align*}
-\min_{t}\{-mt\}  \xrightarrow{\mathrm{Wolfe}} &-\max_{t,z_{1},z_{2}}\{L(m,t,z_{1},z_{2})\} \\
&= \min_{t,z_{1},z_{2}}\{-L(m,t,z_{1},z_{2})\}.
\end{align*}
In summary, for some cost function $C(m)$ and the penalty or regularization term with the ReLU-type function $f(m)$ in \eqref{eq:Relu_min}, the following QUBO formulation is derived:
\begin{align}
&\min_{m}\{C(m) + f(m)\} \nonumber \\
&= \min_{m,t,z_{1},z_{2}}\{C(m) + mt + z_{1}(t+1) - z_{2} t \nonumber \\
& \hspace{20mm} + M \left(-m-z_{1}+z_{2}\right)^2 \},
\label{eq:final_result}
\end{align}
where $ -1 \leq t \leq 0 $, $ z_{1} \geq 0 $ and $ z_{2} \geq 0 $ should be employed by appropriate binary expressions for variables $t$, $z_1$, and $z_2$.
As for the binary expansion, see the Appendix.
Using this final expression, all variables are minimized simultaneously in the annealing methods.

\section{Conclusion}
In the present paper, we proposed a derivation of a two-body Ising representation for a function with a linear part. In addition to the Legendre transformation, the Wolfe dual theorem is used to combine the function with other cost functions within the annealing framework.

As denoted in Secs.~II and III, although the $q$-loss function in \cite{Denchev} has robust characteristics against label noise, the proposed function in the present paper does not have the robustness. 
A derivation of the QUBO formulations for functions with both robust parts and linear parts is a remaining task.
Of course, the necessity of the robust parts is still not clear;
for example, the robust characteristics are not needed for the IMRT problem in \cite{Nazareth2015}.
Actually, in future, we will try to tackle practical optimization problems which need the ReLU-type functions as costs or penalties.
However, it will be also interesting to study differences caused by the quadratic part in the $q$-loss function and the linear part in the ReLU-type function for the label noise problems in the machine learning context \cite{Denchev}. Although these are beyond the scope of the present paper, the derivation technique with the Wolfe theorem could become a key step to try these future studies.

In addition, a sparsity concept is also one of important research topics in machine learning. A typical method to introduce the sparsity is the Lasso proposed by Tibshirani \cite{Tibshirani}. The compressed sensing proposed by Donoho \cite{Donoho} uses the sparsity concept, and it has already been applied to MRI \cite{Lustig} and Networked Data \cite{Haupt}. In the sparsity concept, $\ell_1$ norm-type constraints are used in general. As far as we know, there is no work to derive a two-body Ising expression for $f(m)=|m|$. Note that it is possible to apply the derivation method proposed in the present paper to the $\ell_1$ norm-type functions, and we have already obtained the two-body Ising expression (The paper is in preparation.) 
In addition, discussions for cardinality penalization have already been performed in \cite{Denchev2015}. Although the ReLU-type functions can be used as penalty terms and combined with simple convex cost functions, as in \cite{Nazareth2015}, there is no need to use a convex cost function $C(m)$ in \eqref{eq:final_result}; it is possible to use a non-convex $C(m)$. Considering these non-convex problems, the domain-specific machines with the annealing concepts will give superior capabilities compared with classical computers. We believe that the final formulation and the derivation techniques will be beneficial for various researchers in the communities of annealing methods, ranging from optimization theory to machine learning.



\appendix
\section{Binary Expansion}
Variables used in \eqref{eq:final_result} take continuous values. Therefore, it is necessary to express them by binary expansions for the use in the QUBO formulations. The binary expansions for the $q$-loss function case were described in \cite{Denchev} in detail; please see it too.

When solving problems using \eqref{eq:final_result}, $m$ is often represented by a linear sum like
\begin{align}
\label{eq:m}
m = \sum_{d=1}^{D} w_{d}x_{d},
\end{align}
where $D$ is the dimension of the input space, $w_{d} \in \mathbb{R}$ a coefficient for $d$-th dimension, and $x_{d}\in \mathbb{R}$ the input in $d$-th dimension. Even if this expression is used, \eqref{eq:final_result} remains the two-body Ising form. 

Let $d_{w_{n}}$, $d_{t}$, $d_{z_{1}}$, $d_{z_{2}}$ be bit depths of $w_{n}$, $t$, $z_{1}$, $z_{2}$, respectively.
We denote the discrete variables by $\dot{w}_{n}$, $\dot{t}$, $\dot{z}_{1}$, $\dot{z}_{2}$.
We also define multiplier-offset pairs $(\alpha_{w},\beta_{w})$, $(\alpha_{t},\beta_{t})$, $(\alpha_{z_{1}},\beta_{z_{1}})$, $(\alpha_{z_{2}},\beta_{z_{2}})$ that determine the intervals in which the discrete variables take values.
We perform binary expansions for variables $w_{n}$, $t$, $z_{1}$, $z_{2}$ using a shorthand function $\delta_{\{ w, t, z_{1}, z_{2}\}}(k)=2^{k-1}/(2^{d_{\{ w, t, z_{1}, z_{2}\}}}-1)$:
\begin{align}
w_{n} &\to\dot{w}_{n}=\alpha_{w}\sum_{k=1}^{d_{w}}w_{n,k}\delta_{w}(k) + \beta_{w},\\
t & \to\dot{t}=\alpha_{t}\sum_{k=1}^{d_{t}}t_{k}\delta_{t}(k) + \beta_{t},\\
z_{1} & \to\dot{z}_{1}=\alpha_{z_{1}}\sum_{k=1}^{d_{z_{1}}}z_{1,k}\delta_{z_{1}}(k) + \beta_{z_{1}},\\
z_{2} & \to\dot{z}_{2}=\alpha_{z_{2}}\sum_{k=1}^{d_{z_{2}}}z_{2,k}\delta_{z_{2}}(k) + \beta_{z_{2}}.
\end{align}
The intervals, in which the discrete variables take values, are:
\begin{align}
\dot{w}_{n} & \in [\beta_{w};\alpha_{w} + \beta_{w}], \\
\dot{t} & \in [\beta_{t};\alpha_{t} + \beta_{t}], \\
\dot{z}_{1} & \in [\beta_{z_{1}};\alpha_{z_{1}} + \beta_{z_{1}}], \\
\dot{z}_{2} & \in [\beta_{z_{2}};\alpha_{z_{2}} + \beta_{z_{2}}].
\end{align}
In order to satisfy the constraint condition $-1\leq t \leq 0$, $z_{1}\geq 0$, $z_{2}\geq 0$, we set $\beta_{w}=1-\frac{\alpha_{w}}{2}$, $\alpha_{t}=1$, $\beta_{t}=-1$, $\beta_{z_{1}}=0$, $\beta_{z_{2}}=0$, which gives the intervals:
\begin{align}
\dot{w}_{n} & \in [1-\frac{\alpha_{w}}{2};1+\frac{\alpha_{w}}{2}], \\
\dot{t} & \in [-1;0], \\
\dot{z}_{1} & \in [0;\alpha_{z_{1}}], \\
\dot{z}_{2} & \in [0;\alpha_{z_{2}}].
\end{align}




\end{document}